\documentclass[onecolumn,floatfix,superscriptaddress,showpacs,showkeys,nofootinbib,preprint]{revtex4}

\usepackage{epsfig}
\usepackage{amssymb,latexsym,amsmath}
\newcommand{\eq}[1]{\begin{align} #1 \end{align}}

\begin{document}
\title{ Particle number fluctuations in nuclear collisions\\
 within excluded volume hadron gas model}

\author{M.I. Gorenstein}
\affiliation{ Bogolyubov Institute for Theoretical Physics, Kiev,
Ukraine} \affiliation{Frankfurt  Institute for Advanced Studies,
Frankfurt, Germany}
\author{M. Hauer}
\affiliation{Helmholtz Research  School, University of Frankfurt,
Frankfurt, Germany} \affiliation{University of Cape Town, South
Africa}
\author{D.O. Nikolajenko}
\affiliation{Shevchenko National University, Kiev, Ukraine}
\begin{abstract}
The multiplicity fluctuations are studied in the van der Waals
excluded volume hadron-resonance gas model. The calculations are
done in the grand canonical ensemble within the Boltzmann
statistics approximation. The scaled variances for positive,
negative and all charged hadrons are calculated along the chemical
freeze-out line of nucleus-nucleus collisions at different
collision energies. The multiplicity fluctuations are found to be
suppressed in the van der Waals gas. The numerical calculations
are presented for two values of hard-core hadron radius,
$r=0.3$~fm and 0.5~fm, as well as for the upper limit of the
excluded volume suppression effects.

\end{abstract}



\maketitle
\section{Introduction}
\noindent During last decades statistical models of strong
interactions have served  as an important tool to investigate high
energy nuclear collisions. The main object of the study have been
the mean multiplicities of produced hadrons (see e.g. Refs.
\cite{FOC,FOP,pbm}). Only recently, due to a rapid development of
experimental techniques, first measurements of fluctuations of
particle multiplicities \cite{fluc-mult} and transverse momenta
\cite{fluc-pT} were performed. The study of event-by-event
fluctuations in nucleus-nucleus collisions (see e.g., reviews
\cite{fluc1}) is motivated by expectations of anomalies in the
vicinity of the onset of deconfinement \cite{ood} and in the case
when the expanding system goes through the transition line between
the quark-gluon plasma and the hadron gas \cite{fluc2}.
Furthermore, it is expected that a critical point of strongly
interacting matter may be signaled by a characteristic pattern in
fluctuations \cite{fluc3}.

A convenient measure of the particle number fluctuations is the
scaled variance, $\omega \equiv Var(N)/\langle N\rangle$, defined
in terms of the mean value, $\langle N\rangle$, and variance,
$Var(N)\equiv \langle N^2\rangle - \langle N\rangle^2$. The scaled
variance equals to 1 for the ideal Boltzmann gas in the grand
canonical ensemble (GCE).  The deviations of $\omega$ from unity
in the hadron-resonance gas (HG) come due to Bose and Fermi
statistics, resonance decays (see, e.g., Ref.~\cite{fluc1}), and
exactly enforced conservations laws \cite{CE,res_CE,res_MCE}. 
Previous study \cite{res_MCE} has shown a rather rapid
convergence of the scaled variances to asymptotic values. Finite
system size effects can therefore be ignored for central
collisions of heavy nuclei.

The effect of Bose and Fermi statistics can be seen in the
primordial (before resonance decay) values of the scaled variances
in the GCE (see, e.g., Ref.~\cite{res_CE}). For chemical
freeze-out at low collision energies (small temperature $T$ and
large baryonic chemical potential $\mu_B$) most charged particles
are protons. Fermi statistics then lead to a suppression of
positively charged and all charged particle number fluctuations,
$\omega_+\cong\omega_{ch}=0.98\div 1$.  At high collision energies
(large $T$ and small $\mu_B$) most charged particles are pions.
Thus, Bose statistics dominates and leads to the enhancement of
particle number fluctuations, $\omega_{\pm}\cong
\omega_{ch}=1.05\div 1.06$. These numbers demonstrate that the
effects of quantum statistics are small at the chemical
freeze-out.

Resonance decay in the GCE and CE lead to the enhancement of
particle number fluctuations.
At high collision energies (the largest SPS energy and above) the
GCE calculations give the following final (after resonance decay)
values, $\omega_{\pm}\cong 1.1$ and $\omega_{ch}\cong 1.6$
~\cite{res_CE}.

There is a qualitative difference in the properties of the mean
multiplicity and the scaled variance of multiplicity distribution
in statistical models. In the case of the mean multiplicity
results obtained with the GCE, canonical ensemble (CE), and
micro-canonical ensemble (MCE) approach  each other in the large
volume limit (this reflects the thermodynamical equivalence of
statistical ensembles). It was recently found
\cite{CE,res_CE,res_MCE} that corresponding results for the scaled
variance are different in different ensembles, and, thus, the
scaled variance is sensitive to conservation laws obeyed by a
statistical system. The differences are preserved  in the
thermodynamic limit. The global conservation laws imposed on each
microscopic state of the statistical system lead to the
suppression of particle number fluctuations.  At large $T$ and
small $\mu_B$ (high collision energies) the final state scaled
variances behave in the CE as, $\omega_{\pm}\cong 0.8$  and
$\omega_{ch}\cong 1.6$ \cite{res_CE}, and in the MCE as, $\omega_{\pm}\cong 0.3$
and $\omega_{ch}\cong 0.6$ ~\cite{res_MCE}.

It has been found that the fluctuations in the number of nucleon
participants give a large contribution to hadron multiplicity
fluctuations in peripheral A+A collisions. Thus ideally a
comparison between data and predictions of statistical models should be
performed for results which correspond to collisions with a fixed number of
nucleon participants. This can approximately be achieved by
selecting only the most central A+A collisions (see
Ref.~\cite{Voka,MGMG} for details). The conditions for the centrality
selection in the study of fluctuations should thus be much more
stringent than in the multiplicity measurements. To minimize the
contributions from the participant number fluctuations, one should restrict the
analysis to the 1\%  most central A+A collisions selected by the number of
projectile spectators (see Refs.~\cite{res_MCE,NA49}). In this sample
both the fluctuations of the number of participating nucleons and the
resulting enhancement of multiplicity fluctuations should not exceed  a few \%.

Usually the ideal HG gas is used for the particle multiplicity
calculations.
The extension of the ideal gas picture based on the van der Waals
(VDW) excluded volume procedure have been suggested  in
Refs.~\cite{vdw,vdw1,crit} to phenomenologically take into account
repulsive interactions between hadrons.
This leads to the suppression of
hadron number densities and energy density.
In the present paper we study the effect of the VDW excluded
volume on the multiplicity fluctuations.  The particle number
ratios are independent of the proper volume parameter, if it is
the same for all hadron species. Thus, a rescaling of the total
volume $V$ leads to exactly the same hadron yields as those in the
ideal HG. In the present paper it is demonstrated that the
multiplicity fluctuations are suppressed in the VDW gas. This
suppression is qualitatively different from that of
particle yields. In contrast to average multiplicities, the
suppression of multiplicity fluctuations can not be removed by
rescaling of the total volume of the system.

In the present study we restrict ourself to the GCE calculations
within the Boltzmann statistics approximation.
The paper is organized as follows. In Section II the ideal HG model
is presented.  In Section III the one-component VDW gas is used to
calculate the average multiplicity and the scaled variance. These
results are extended to the multi-component VDW gas in Section IV.
Both primordial and final state (after resonance decay) hadron
yields and their fluctuations are considered. In Section V the
scaled variance of negative, positive and all charged hadrons is
calculated along the chemical freeze-out line. This gives the VDW
HG predictions for fluctuations of hadron multiplicities in
central A+A collisions at different collision
energies from SIS to LHC.  A summary, presented in Section VI,
closes the paper.

\section{Ideal Hadron-resonance Gas}
\noindent The partition function of the ideal Boltzmann HG in the
GCE has the form:
\eq{\label{Z-id} Z_{id}~=~\exp\left[\sum_j V~\phi_j ~
\right]~,
}
where:
\begin{align}\label{phi-j}
\phi_j~  \equiv
~\lambda_j~\frac{g_j}{2\pi^2}~~\int_0^{\infty}k^2dk~
\exp\left[-~\frac{(k^2~+~m_j^2)^{1/2}}{T} \right]
=~\lambda_j~\frac{g_j}{2\pi^2}~m_j^2~T~K_2\left(\frac{m_j}{T}\right)~.
\end{align}
In Eqs.~(\ref{Z-id},\ref{phi-j}),
$\lambda_j=\exp\left(\mu_j/T\right)$ is the particle specific
fugacity, $g_j$ the degeneracy of $j$th particle species, $m_j$
the particle mass,
and $\mu_j = b_j\mu_B + s_j\mu_S  + q_j\mu_Q $ the chemical
potential due to the $j$th  particle  internal quantum numbers
$\left(b_j,s_j,q_j\right)$. The $j$-summation in
Eq.~(\ref{Z-id}) is taken over all hadron species.
Global chemical potentials ($\mu_B$, $\mu_S$, $\mu_Q$) and
temperature $T$ regulate the systems average charges (baryonic,
strangeness, electrical) and energy densities. Finally, $V$ is the
volume of the system and $K_2$ is the modified Hankel function.

\subsection{Primordial Fluctuations}
\noindent In the GCE the first two moments  of the multiplicity
distribution of $j$th particle  are:
\begin{align}\label{Nid}
\langle N_j\rangle~&=~\frac{1}{Z_{id}} \left( \lambda_j
  \frac{\partial}{\partial \lambda_j} 
\right)~ Z_{id} ~=~ V~\phi_j~,\\
\langle N_j^2 \rangle  ~&=~\frac{1}{Z_{id}} \left( \lambda_j
\frac{\partial}{\partial \lambda_j} \right)^2 ~ Z_{id} ~=~ \left(
V \phi_j \right)^2 ~+~V \phi_j~.
\end{align}
Thus, $\phi_j$ is the $j$th hadron number density, $n_j^{id}$.
The variance of the $j$th particle number distribution is:
\begin{align}\label{DeltaNj}
\langle \left( \Delta N_j \right)^2 \rangle ~\equiv~ \langle
\left(N_j~-~\langle N_j\rangle \right) \rangle^2~=~\langle N_j^2 \rangle -
\langle N_j \rangle^2 ~=~ \langle N_j \rangle~,
\end{align}
and the scaled variance equals unity,
\begin{align}\label{omegaid}
 \omega_j ~\equiv~\frac{\langle \left( \Delta N_j \right)^2 \rangle}
{\langle N_j\rangle}~=~1~.
\end{align}
For the average multiplicity and  variance of, for example, positively
charged primordial hadrons (i.e. before resonance decay) one
obtains:
\begin{align}\label{positive}
\langle N_+\rangle ~= ~\sum_{j,~q_j>0}\langle N_j \rangle ~,\qquad
~ \qquad ~ \langle\left( \Delta N_+ \right)^2 \rangle~ =~
\sum_{j,~q_j>0} ~\sum_{k,~q_k>0} ~\langle \Delta N_j \Delta N_k
\rangle~.
\end{align}
The results for the negatively charged or all charged hadrons are
similar to Eq.~(\ref{positive}) with summation over negative
charges, $q_j<0,~q_k<0$, or all non-zero charges, $q_j\neq
0,~q_k\neq 0$, respectively. For $j\neq k$, one finds,
\begin{align}\label{jneqk}
\langle N_j  N_k \rangle ~=~ \frac{1}{Z_{id}} \left[\left(
\lambda_j
  \frac{\partial}{\partial \lambda_j} \right)~ \left( \lambda_k
  \frac{\partial}{\partial \lambda_k} \right)\right]~ Z_{id} ~=~ \langle
N_j \rangle ~\langle N_k \rangle~,
\end{align}
and from Eqs.~(\ref{DeltaNj}) and (\ref{jneqk}) it then follows,
\begin{align}\label{jkcorr}
\langle \Delta N_j \Delta N_k \rangle ~=~ \langle N_j  N_k \rangle
- \langle N_j \rangle ~\langle N_k \rangle ~=~ \delta_{jk}~
\langle N_j \rangle~.
\end{align}
From Eq.~(\ref{jkcorr}) one concludes that in the ideal HG the
correlations between different particle species are absent in the
GCE. Any particle number distribution in the ideal Boltzmann gas
is  Poissonian  (for large average multiplicity this is equivalent
to a Gaussian) and its scaled variance therefore equals unity. For
example, the scaled variance for positively charged hadrons is,
\begin{align}\label{omegaeq1}
\omega_+ ~=~ \frac{\langle \left( \Delta N_+ \right)^2
\rangle}{\langle N_+ \rangle}  ~=~ 1~.
\end{align}
This is also valid
for negatively charged and all charged
hadrons.

\subsection{Final State Fluctuations}
\noindent Final state yields and (co)variance have simple forms in
the GCE \cite{Koch}:
\begin{align}\label{av-id}
\langle N_j \rangle ~&=~ \langle N_j^* \rangle ~+~ \sum_R ~\langle N_R \rangle
~\langle n_j \rangle_R~, \\
\langle \Delta N_j \Delta N_k \rangle ~&=~ \langle\left(\Delta N_j^*
\right)^2\rangle 
~\delta_{jk} ~+~\sum_{R} ~\langle (\Delta N_{R})^2 ~\rangle
\langle n_j \rangle_R\langle n_k \rangle_R~+~ \sum_{R} ~\langle
N_{R} ~\rangle \langle \Delta n_j \Delta n_k \rangle_R~,
\label{corr-id}
\end{align}
where $N_j^*$ is primordial $j$th hadron multiplicity, and:
\begin{align}
\langle n_j \rangle_R ~&=~ \sum_r ~b_r^R ~n_{j,R}^r ~,\\
 \langle \Delta n_j \Delta n_k \rangle_R~&=~ \langle n_j n_k
\rangle_R ~-\langle n_j\rangle_R \langle n_k\rangle_R ~=~ \sum_r
~b_r^R ~n_{j,R}^r ~n_{k,R}^r~-~\langle n_j \rangle_R\langle n_k \rangle_R~,
\end{align}
$b_r^R$ is the branching ratio of the $r$th decay channel  of
$R$th resonance  with normalization condition $\sum_r b_r^R=1$,
while $n_{j,R}^r$ and $n_{k,R}^r$ are the respective
multiplicities of particles of species $j$ and $k$ in $r$th decay
channel of $R$th resonance.
Generally, one finds enhancement of fluctuations due to
resonance decay which leads to $\omega_j > 1$ (valid in GCE and CE, see, e.g.,
Ref.~\cite{res_CE}). The first term in the right hand side of
Eq.~(\ref{corr-id}) corresponds to fluctuation of $R$th resonance
multiplicity and the second term appears due to the probabilistic
character of resonance decay. For the Boltzmann gas discussed in
this paper, it follows, $\langle (\Delta N_R)^2\rangle=\langle
N_R\rangle$, $\langle (\Delta N_j)^2\rangle=\langle
N_j^*\rangle$, and Eq.~(\ref{corr-id}) simplifies to:
\eq{
\langle \Delta N_j \Delta N_k \rangle ~=~ \langle N_j^* \rangle
~\delta_{jk} ~+~\sum_{R} ~\langle N_{R} ~\rangle  \langle  n_j n_k
\rangle_R~. \label{corr-id-B}
}
From Eqs.~(\ref{av-id}) and (\ref{corr-id-B}) it immediately
follows,
\eq{ \label{omega-id-1}
\omega_j~=~\frac{\langle N_j^* \rangle ~ ~+~\sum_{R} ~\langle
N_{R} ~\rangle  \langle  n_j^2 \rangle_R}{\langle N_j^* \rangle ~
~+~\sum_{R} ~\langle N_{R} ~\rangle  \langle  n_j \rangle_R}~.
}
Thus, $\omega_j\ge 1$ as a
consequence of the inequality $(n_{j,R}^r)^2\ge n_{j,R}^r$~, i.e.
the scaled variance of $j$th hadron exceeds 1 due to the presence of
 resonances  decaying  into more than one $j$th hadron.
Using Eqs.~(\ref{av-id}) and (\ref{corr-id-B}) for average yields
and correlators one finds from (\ref{positive}),
\eq{
\omega_+~=~\frac{\sum_{j,~q_j>o}~[~\langle N_j^* \rangle ~
~+~\sum_{k,~q_k>0}\sum_{R} ~\langle N_{R} ~\rangle
\langle n_jn_k \rangle_R~]}{\sum_{j,~q_j>o}~[~\langle N_j^* \rangle ~
~+~\sum_{R} ~\langle N_{R} ~\rangle  \langle  n_j \rangle_R~]}~,
}
and similar expressions for final state $\omega_-$  and
$\omega_{ch}$ with the summation over $q_j<0,~q_k<0$ and $q_j\neq
0,~q_k\neq 0$, respectively.

\section{One-Component VDW Gas  }
\noindent The VDW excluded volume procedure gives the GCE
partition function of one component Boltzmann gas \cite{vdw}:
\eq{\label{Zvdw}
Z(V,T,\mu_j)~=~\sum_{N=0}^{\infty}~\frac{1}{N!}~\left[
(V~-~v_jN)~\phi_j~\right]^N
~\theta(V-v_jN)~,
}
where $v_j$ is the proper volume of the $j$th particle. For the
`hard sphere' particles with radius $r_j$ the volume parameter
$v_j$ equals the `hard-core particle volume', $4\pi r_j^3/3$,
multiplied by a factor of 4. A Laplace transform of
Eq.~(\ref{Zvdw}) reads:
\eq{
 \label{Zvdw1}
& \widehat{Z}(s,T,\mu_j)~\equiv~
\int_0^{\infty}dV\exp(-sV)~Z(V,T,\mu_j)\nonumber ~
=~\sum_{N=0}^{\infty}~\frac{
\phi_j^N}{N!}~
\int_{v_jN}^{\infty}dV\exp(-sV)~\left(V~-~v_jN\right)^N~\nonumber \\
 &=~\left[~s~-~\exp(-v_js)~\phi_j\right]^{-1}~.
}
The system pressure is defined by the
pole-singularity, $s^*$, of the function, $\widehat{Z}$
(\ref{Zvdw1}),
\eq{p~\equiv ~\lim_{V\rightarrow\infty}~ T~\frac{\ln
Z}{V}~=~T~s^*~,
}
and, thus, satisfies the following equation,
\begin{align}\label{pvdw}
p~=~
\exp\left(- ~\frac{v_j~ p}{T}\right)~T~\phi_j~.
\end{align}
For $v_j=0$, Eq.(\ref{pvdw}) is evidently reduced to the ideal gas
result, $p_{id}=T\phi_j$.
The average particle number in the VDW gas equals to:
\begin{align}\label{Nav-vdw}
\langle N_j\rangle ~=~ \frac{1}{Z(V,T,\mu_j)} \left( \lambda_j
  \frac{\partial}{\partial \lambda_j} \right)~Z(V,T,\mu_j)
~=\frac{V}{T}~\lambda_j~\frac{\partial p}{\partial\lambda_j}~.
\end{align}
To find the particle number fluctuations one needs to calculate,
\begin{align}\label{N2}
\langle N_j^2\rangle ~=~
\frac{1}{Z}~\left(\lambda_j\frac{\partial}{\partial \lambda_j}
\right)^2~ Z ~=~ \frac{V}{T}~\lambda_j~\frac{\partial
p}{\partial\lambda_j}~ + ~\left(\frac{V}{T} \right)^2
~\lambda_j^2~\left(\frac{\partial p}{\partial\lambda_j}\right)^2
~+~\frac{V}{T}~\lambda_j^2~ \frac{\partial^2
p}{\partial\lambda_j^2}~.
 \end{align}
The variance is therefore:
\begin{align}\label{var_vdw}
\langle \left( \Delta N_j \right)^2 \rangle ~=~  \frac{V}{T} ~
\left[ \lambda_j~ \frac{\partial p}{\partial \lambda_j}~ +~
 \lambda_j^2~ \frac{\partial^2 p}{\partial\lambda_j^2}\right]~.
\end{align}
Thus, one can write the scaled variance as the following,
\begin{align}\label{omega_j_vdw}
\omega_j ~= ~\frac{\langle \left( \Delta N_j \right)^2
\rangle}{\langle N_j
  \rangle}~ =~ 1~+~\lambda_j~ \frac{\partial^2
    p}{\partial\lambda_j^2}~ \times ~\left(\frac{\partial
    p}{\partial\lambda_j}\right)^{-1}~.
\end{align}
From Eq.~(\ref{pvdw}) one finds:
\begin{align}\label{p1lambda}
\lambda_j~\frac{\partial p}{\partial\lambda_j}~
=~T~x_j~-~\lambda_j~v_j~x_j~\frac{\partial p}{\partial\lambda_j}~,
\end{align}
where  the notation,
\eq{ \label{x-j}
x_j~\equiv~\exp\left(-~\frac{v_j~p}{T}\right)~\phi_j~,
}
is introduced to simplify the formulas below. From
Eq.~(\ref{p1lambda}) it follows,
\begin{align}\label{p1lambda1}
\lambda_j  ~\frac{\partial p}{\partial\lambda_j}~=~
\frac{T~x_j}{1~+~v_jx_j}~,
\end{align}
and one finds:
\begin{align}\label{Nvdw}
\langle N_j\rangle~=~\frac{V~x_j}{1~+~v_j~ x_j}~.
\end{align}
From Eqs.~(\ref{pvdw},\ref{x-j},\ref{Nvdw}) one finds the VDW equation of
state, familiar from textbooks,
\eq{
p~\left[V~-~v_j\langle N_j\rangle\right]~=~\langle N_j\rangle~T~.
}
At small particle density, $v_jx_j\ll 1$, one finds from
Eq.~(\ref{Nvdw}),
\begin{align}\label{n1vdw}
\langle N_j\rangle ~\cong V~\phi_j~\left(1~-~2v_j\phi_j\right) ~=~\langle
N_j^{id}\rangle~\left(1~-~2v_jn_j^{id}\right)~,
\end{align}
where $n_j^{id}=\phi_j$ is the ideal gas particle
number density.
In the opposite limiting
case of high particle density, $v_jx_j\gg 1$, one obtains,
\begin{align}\label{n2vdw}
N_j~\cong ~\frac{V}{v_j}~,
\end{align}
thus, the value of $1/v_j$ is the upper limit for the particle
number density in the VDW gas (\ref{Zvdw}).
From Eq.~(\ref{p1lambda}) and (\ref{p1lambda1}) it follows,
\begin{align}\label{p2lambda1}
\lambda_j^2~\frac{\partial^2p}{\partial \lambda_j^2}~ =~
-~\frac{Tv_jx_j^2}{(1~+~v_jx_j)^2}~\left[2~-~
\frac{v_jx_j}{1~+~v_jx_j}\right]~.
\end{align}
From Eqs.~(\ref{omega_j_vdw},\ref{p1lambda1},\ref{p2lambda1}) one
then obtains,
\begin{align}\label{omegavdw}
\omega_j ~
=~\left(1~+~v_j~x_j\right)^{-2}~.
\end{align}
From Eq.~(\ref{omegavdw}) it follows that the scaled variance in
the VDW gas is always smaller than the ideal gas value
(\ref{omegaid}) $\omega_j = 1$, i.e. the non-zero proper volume
$v_j$ suppresses the particle number fluctuations. At small
particle number densities, $v_jx_j \ll 1$, the particle number
fluctuations in the VDW gas are approximately equal to those in an
ideal gas (\ref{omegaid}). The Eq.~(\ref{omegavdw}) gives,
 $\omega_j~\cong~1~-~2v_j\phi_j$~.
On the other hand, at $v_jx_j\gg 1$ the
scaled variance of the VDW gas goes to zero as $\omega_j\cong
(v_jx_j)^{-2}$.  In this limit the particle number density is
close to its lower limit, $n_j\cong 1/v_j$. Thus, there is no 'free space'
for particle number fluctuations.

\section{ Multi-Component VDW Gas}
\noindent
The partition function of the multi-component VDW gas
equals to \cite{vdw,vdw1,crit}:
\begin{align}\label{Zvdw_HRG}
Z~=~\prod_{j=1}~\sum_{N_j=0}^{\infty}~\frac{1}{N_j!}~\left[(V~-~
\sum_i v_iN_i)~
\phi_j\right]^{N_j}
~\theta(V~-~\sum_i v_iN_i)~.
\end{align}
Similar to Eq.~(\ref{pvdw}) one finds the equation for the
pressure function in the multi-component VDW gas,
\begin{align}\label{p-mvdw}
p~=~ \sum_j~~T~x_j~,
\end{align}
where $x_j$ is given by Eq.~(\ref{x-j}).

\subsection{Primordial Fluctuations}
\noindent The average multiplicity of $j$th particle is:
\begin{align}\label{Nj-mvdw}
\langle N_j\rangle ~=~\lambda_j~\frac{\partial}{\partial
\lambda_j} \ln Z~=~\frac{V}{T} ~\lambda_j~\frac{\partial
p}{\partial\lambda_j}~.
\end{align}
Similar to Eq.~(\ref{var_vdw}) one obtains:
\begin{align}\label{NjVDW}
\langle  \Delta N_j \Delta N_k \rangle ~=~ \frac{V}{T} \left[
  \delta_{jk} ~\lambda_j  \frac{\partial p}{\partial
      \lambda_j}~ + ~\lambda_j\lambda_k~ \frac{\partial^2p}{\partial
      \lambda_j \partial \lambda_k}  \right]~.
\end{align}
Calculating the derivative 
from Eq.~(\ref{p-mvdw}),
\begin{align}\label{dp}
\lambda_j~\frac{\partial p}{\partial\lambda_j}~=~
\frac{T~x_j}{1~+~\sum_i~v_i~x_i}~,
\end{align}
and using Eq.~(\ref{Nj-mvdw}) one finds:
\begin{align}\label{Nmvdw}
\langle N_j\rangle~=~\frac{V~x_j}{1~+~\sum_i~v_i~x_i}~,~~~~~
\end{align}
which extends  Eq.~(\ref{Nvdw}) to the multi-component VDW
gas. Using Eq.(\ref{dp}) for the first derivatives of the pressure one
finds:
\begin{equation}\label{pjk}
 \lambda_j\lambda_k~\frac{\partial^2 p}{\partial \lambda_j \partial \lambda_k }
~ = ~ - \frac{T~ x_j
  x_k}{\left(1+ \sum_i x_i v_i \right)^2} \left[v_j + v_k - \frac{\sum_i x_i
    v_i^2}{\left(1+ \sum_i x_i v_i \right) } \right]~.
\end{equation}
The Eq.~(\ref{pjk}) for $j=k$ leads to the following result for $\omega_j$,
\begin{align}\label{omega-j-vdw}
\omega_j ~=~ 1~ +~
\lambda_j~ \frac{\partial^2 p}{\partial \lambda_j^2 }  ~ \times
~\left(   \frac{\partial p}{\partial
  \lambda_j} \right)^{-1}
  ~=~1~-~\frac{x_j}{1~+~\sum_iv_ix_i}~~\left(2v_j~-~\frac{\sum_i
v_i^2 x_i} {1~+~\sum_i v_ix_i}\right)~.
\end{align}
The Eq.~(\ref{omega-j-vdw}) is reduced to Eq.~(\ref{omegavdw})
for the one-component VDW gas. For an `almost` ideal gas when all
proper volumes come to zero, $v_i \rightarrow 0$, one finds the
ideal gas results, $\langle N_j \rangle \cong V\phi_j$ and
$\omega_j \cong 1$, from Eqs.~(\ref{Nmvdw}) and
(\ref{omega-j-vdw}), respectively. The above equations are
simplified if the proper volumes of different hadron species are
equal to each other, $v_i=v$. The scaled variance
(\ref{omega-j-vdw}) becomes then equal to:
\begin{align}\label{omega-j-vdw-a}
\omega_j
  ~=~1~-~v~\frac{x_j~(2~+~v~\sum_ix_i)}
 {(1~+~v~\sum_i x_i)^2}~.
\end{align}
The Eq.~(\ref{omega-j-vdw-a}) demonstrates a suppression of the
particle number fluctuations for all hadron species ($\omega_j <
1$).  At small total particle number density, $v\sum_ix_i\ll 1$,
it behaves as, $\omega_j\cong 1- 2v\phi_j$, similar to the
one-component VDW gas. In the opposite limiting case,
$v\sum_ix_i\gg 1$, one finds, $\omega_j\cong 1-
\phi_j/\sum_i\phi_i=1-n_j^{id}/n_{tot}^{id}$. In a dense VDW gas
the particle number fluctuations are suppressed, and the
suppression for $j$th species is proportional to the ratio of the
ideal gas $j$th particle number density, $n_j^{id}$, to the total
particle number density, $n_{tot}^{id}=\sum_in_i^{id}$. The
largest suppression takes place for the total multiplicity,
$\omega_{tot}\rightarrow 0$ at $v\sum_ix_i\rightarrow\infty$. This
is in an agreement with the behavior of the scaled variance in
one-component VDW gas at high particle number density.

For the correlators (\ref{NjVDW}) in the VDW gas one obtains,
\eq{\label{VDW-corr}
\langle  \Delta N_j \Delta N_k \rangle ~=~  \langle N_j\rangle
~\left[~\delta_{jk}~ -~v~\frac{x_k~(2~+~v\sum_ix_i)}{\left(
1~+~v\sum_ix_i \right)^2}~\right]~.
}
In contrast to the ideal HG result  (\ref{jkcorr}), the excluded
volume effects lead to the (anti)correlation between different
particle species, $j\neq k$, in the VDW gas seen from
Eq.~(\ref{VDW-corr}). The physical origin of these
anticorrelations is rather clear. The `large' number of $j$th
particles, $\Delta N_j>0$, reduces the available free space for
$k$th particles. This makes preferable to have a `small' number of
$k$th particles, $\Delta N_k<0$, thus, leading to  $\langle
\Delta N_j \Delta N_k \rangle<0$.

Using Eq.~(\ref{VDW-corr}) one finds the primordial fluctuations
of the positively charged hadrons,
\begin{align} \label{omega-plus-VDW}
\omega_+ ~&=~ \frac{\sum_{j,~q_j > 0}  \sum_{k,~q_k >0}~\langle
\Delta N_j \Delta N_k \rangle } {\sum_{i,~q_i>0} \langle
N_i\rangle}
\nonumber \\
&= ~1~ -~ v~ \left[ \sum_{j,q_j > 0} \sum_{k,q_k >
    0 } \frac{x_j~ x_k \left( 2~ +~
      v~\sum_i x_i \right)}{\left(1~+~
       v \sum_i x_i \right)^2 }\right]    ~\times ~
      \left(\sum_{j,q_j>0}~ x_j\right)^{-1}~.
\end{align}
and similar expressions for $\omega_-$  and
$\omega_{ch}$ with the summation over $q_j<0,~q_k<0$ and $q_j\neq
0,~q_k\neq 0$, respectively.

\subsection{Final State Fluctuations}
\noindent To take  into account resonance decay we follow the
procedure from Refs.~\cite{res_CE,res_MCE}. The correlators for
the final hadrons are expressed as,
\begin{align}
  \langle \Delta N_j\,\Delta N_k\rangle
  ~&=\; \langle\Delta N_j^* \Delta N_k^*\rangle
  \;+\; \sum_R \langle N_R\rangle\; \langle n_{j} n_{k}\rangle_R
  \;+\; \sum_R \langle\Delta N_j^*\; \Delta N_R\rangle\; \langle n_{k}\rangle_R
  \; \nonumber
  \\
  &+\; \sum_R  \langle\Delta N_k^*\;\Delta N_R\rangle\; \langle n_{j}\rangle_R
  \;+\; \sum_{R\neq R'} \langle\Delta N_R\;\Delta N_{R'}\rangle
  \; \langle n_{j}\rangle_R\;
       \langle n_{k}\rangle_{R^{'}}\;.\label{corr-VDW}
\end{align}
where all particle-particle, $\langle\Delta N_j^* \Delta
N_k^*\rangle$, particle-resonance, $\langle\Delta N_j^* \Delta
N_R\rangle$, and resonance-resonance, $ \langle\Delta N_R\Delta
N_{R'}\rangle$, correlators in Eq.~(\ref{corr-VDW}) are calculated
according to Eq.~(\ref{VDW-corr}). Note the essential difference
between the correlators in the ideal gas (\ref{corr-id-B}) and
those in the VDW gas (\ref{corr-VDW}). The new terms in
Eq.~(\ref{corr-VDW}) come from the anticorrelations of different
stable hadron and resonance species in the VDW gas according to
Eq.~(\ref{VDW-corr}).

\section{Scaled Variance along Chemical Freeze-out Line}
\noindent  In this section we present the results of the VDW HG
for the scaled variances along the chemical freeze-out line in
central A+A collisions for the whole energy range from SIS to LHC.
The procedure to define the chemical freeze-out line is
essentially the same as in Refs.~\cite{res_CE,res_MCE}. The values
of $T$ and $\mu_B$ at the chemical freeze-out at different
collision energies are presented in Table I. They are almost
identical to those values in Fig.~1 and Table~I of
Ref.~\cite{res_CE}. The only tiny difference comes because of the
Boltzmann statistics approximation used in the present paper. Note
that the conditions for average energy per particle, $\langle
E\rangle/\langle N\rangle =1$~GeV \cite{Cl-Red},  zero value of
the net total strangeness, $S=0$, and the charge to baryon ratio,
$Q/B = 0.4$, remain the same as in Refs.~\cite{res_CE,res_MCE}.
This procedure is possible  because all particle ratios and energy
to particle ratio in the VDW HG gas (with the same hard-core
volume $v_j=v$ for all hadron species) remain unchanged in a
comparison to the ideal HG, $n_j/n_k=n_j^{id}/n_k^{id}$ and
$\varepsilon/n_{tot}=\varepsilon^{id}/n_{tot}^{id}$. The
dependence of $\mu_B$ on the collision energy is parameterized as
\cite{FOC},
$\mu_B \left( \sqrt{s_{NN}} \right) =1.308~\mbox{GeV}~(1+
0.273~ \sqrt{s_{NN}})^{-1}~,$
where the c.m. nucleon-nucleon collision energy, $\sqrt{s_{NN}}$,
is taken in GeV units. The strangeness saturation factor, $\gamma_S$,
is  parameterized as \cite{FOP},
$ \gamma_S~ =~ 1 - 0.396~ \exp \left( - ~1.23~ T/\mu_B \right)$.
Both these relations are the same as in Refs.~\cite{res_CE,res_MCE}.

\begin{table}[h!]
\begin{center}
\begin{tabular}{||c||c|c|c||c|c||c|c|c||}\hline
 $\sqrt{s_{NN}}$& $T$ &$\mu_B$& $\gamma_S$ &
   \multicolumn{2}{c||}{$R=\varepsilon/\varepsilon^{id}$} 
   &   \multicolumn{3}{c||}{$n_i/n_{tot}$}\\ [0.5ex]
\hline [ GeV ] &[ MeV ] &[ MeV ] & & $r=\!0.3$fm & $r=\!0.5$fm
 & $n_+/n_{tot}$ & $n_-/n_{tot}$& $n_{ch}/n_{tot}$ \\
[0.5ex] \hline\hline
$ 2.32 $ & 64.3  & 800.8 & 0.64  & 0.944& 0.788 & 0.413 & 0.054 & 0.467 \\
$ 4.86 $ & 116.5 & 562.2 & 0.69  & 0.870& 0.603 & 0.388 & 0.185 & 0.573 \\
$ 6.27 $ & 128.5 & 482.4 & 0.72  & 0.844& 0.552 & 0.371 & 0.207 & 0.577 \\
$ 7.62 $ & 136.1 & 424.6 & 0.74  & 0.825& 0.519 & 0.358 & 0.218 & 0.576 \\
$ 8.77 $ & 140.6 & 385.4 & 0.75  & 0.812& 0.498 & 0.349 & 0.225 & 0.574 \\
$ 12.3 $ & 149.0 & 300.1 & 0.79  & 0.786& 0.459 & 0.331 & 0.237 & 0.568 \\
$ 17.3 $ & 154.4 & 228.6 & 0.83  & 0.766& 0.432 & 0.316 & 0.245 & 0.561 \\
$ 62.4 $ & 160.6 & 72.7  & 0.98  & 0.738& 0.397 & 0.285 & 0.263 & 0.548 \\
$ 130  $ & 161.0 & 35.8  & 1.0  & 0.735& 0.393  & 0.278 & 0.268 & 0.546 \\
$ 200  $ & 161.1 & 23.5  & 1.0  & 0.735& 0.393  & 0.277 & 0.269 & 0.546 \\
$ 5500 $ & 161.2 & 0.9   & 1.0  & 0.735& 0.393  & 0.273 & 0.273 & 0.546 \\
\hline
\end{tabular}
\caption{The chemical freeze-out parameters $T$, $\mu_B$, and
$\gamma_S$  in central A+A collisions are presented at different
c.m. energies $\sqrt{s_{NN}}$. The calculations are done for the
HG with Boltzmann statistics in the GCE. The excluded volume
parameter, $v=16\pi r^3/3$, is taken to be the same for all hadron
species. The VDW suppression factor,
$R=\exp(-vp/T)[1+v\sum_ix_i]$, for the VDW energy density and
particle number densities, and the ratios $n_i/n_{tot}=n_i^{id}/n_{tot}^{id}$
(see also Fig.~2, right) are presented.
\label{Table1}}
\end{center}
\end{table}

\begin{figure}[ht!]
\begin{center}
 \epsfig{file=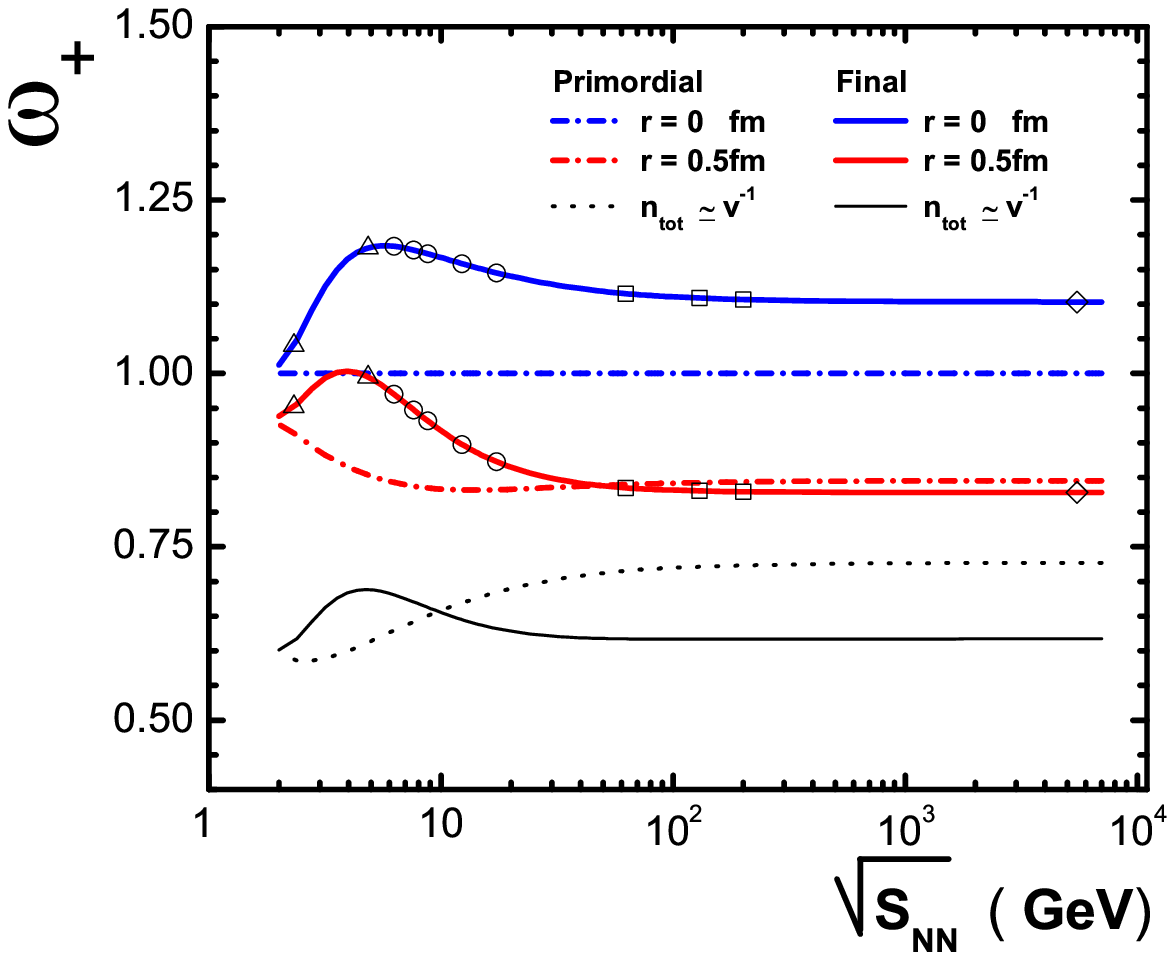,height=6.7cm}
 \epsfig{file=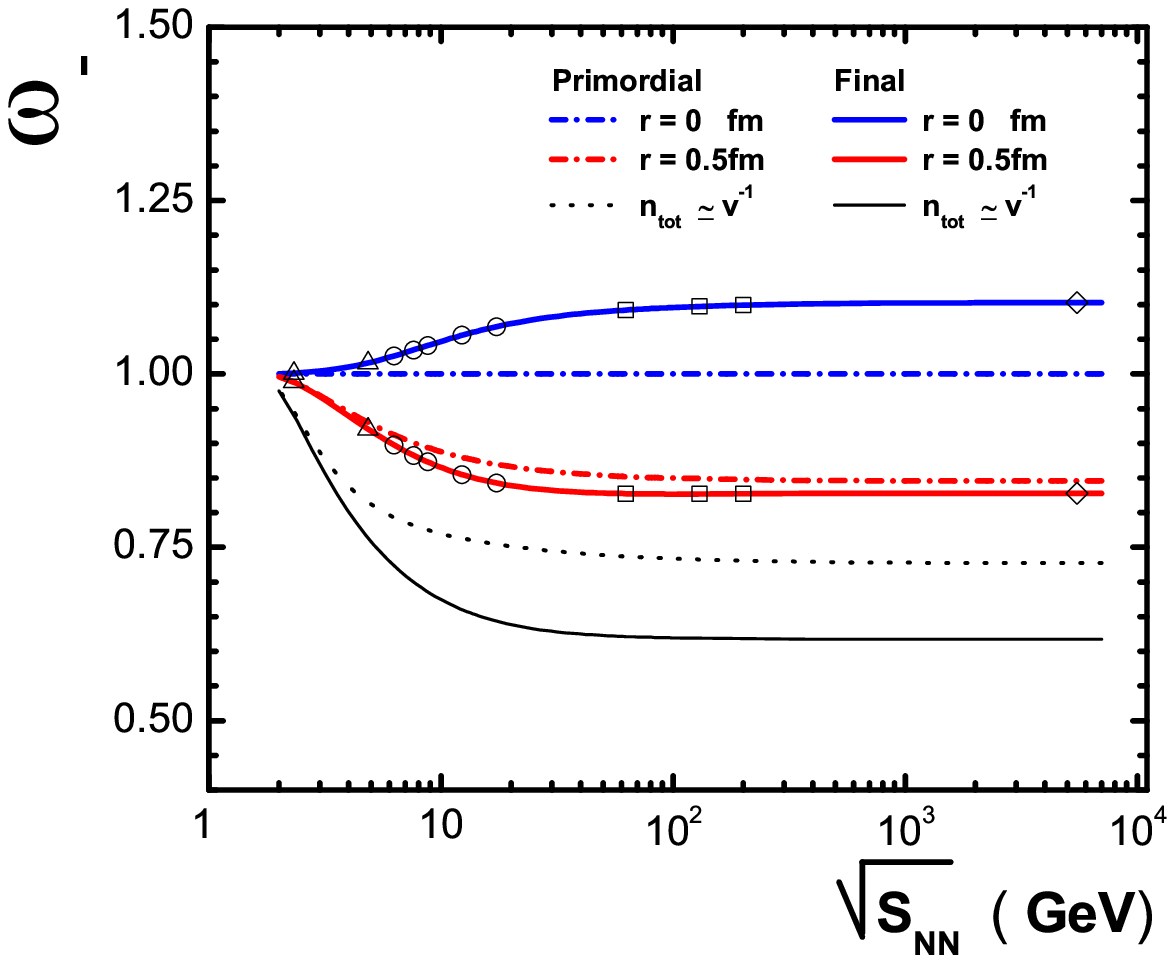,height=6.7cm}
 \caption{The scaled variances for positively ({\it left})
and negatively ({\it right}) charged particles along the chemical
freeze-out line for central A+A collisions (see also Tables II and III). 
The calculations are
done for the HG with Boltzmann statistics in the GCE. Symbols at
the lines correspond to the specific collision energies pointed
out in Table I. The dash-dotted lines correspond to the primordial
and solid lines to final state values. The pair of upper lines
show the ideal HG results, the pair of middle lines the VDW HG
with hard-core radius $r=0.5$~fm. The lowest pair of lines
corresponds to largest possible VDW suppression for the primordial
(dotted) and final (solid) scaled variance, respectively. This
largest VDW suppression happens at $n_{tot}\cong 1/v$. }
\label{omega_pos}
\end{center}
\end{figure}

\begin{table}[h!]
\begin{center}
\begin{tabular}{||c||c|c|c|c||c|c|c|c||}\hline
 $\sqrt{s_{NN}}$& \multicolumn{4}{c||}{ $\;\omega_{+}$~~~Primordial} & \multicolumn{4}{c||}{
  $\;\omega_{+}$~~~Final} \\ [0.5ex]
\hline [ GeV ] &$r=0$ & $r=\!0.3$fm &
$r=\!0.5$fm&$n_{tot}\cong v^{-1}$ &$~r=0~$ & $r=\!0.3$fm & $r=\!0.5$fm & $n_{tot}\cong v^{-1}$ \\
[0.5ex] \hline\hline
$ 2.32 $ &  1  & 0.977 & 0.915 & 0.587 & 1.040 & 1.017 & 0.952 & 0.613  \\
$ 4.86 $ &  1  & 0.951 & 0.853 & 0.612 & 1.181 & 1.118 & 0.995 & 0.688  \\
$ 6.27 $ &  1  & 0.943 & 0.843 & 0.629 & 1.183 & 1.106 & 0.970 & 0.681  \\
$ 7.62 $ &  1  & 0.939 & 0.837 & 0.642 & 1.178 & 1.091 & 0.948 & 0.671  \\
$ 8.77 $ &  1  & 0.936 & 0.835 & 0.651 & 1.173 & 1.079 & 0.932 & 0.663  \\
$ 12.3 $ &  1  & 0.931 & 0.832 & 0.669 & 1.158 & 1.051 & 0.898 & 0.645  \\
$ 17.3 $ &  1  & 0.928 & 0.832 & 0.684 & 1.145 & 1.029 & 0.873 & 0.632  \\
$ 62.4 $ &  1  & 0.928 & 0.840 & 0.715 & 1.115 & 0.989 & 0.835 & 0.617  \\
$ 130  $ &  1  & 0.929 & 0.843 & 0.722 & 1.109 & 0.983 & 0.830 & 0.617  \\
$ 200  $ &  1  & 0.929 & 0.844 & 0.723 & 1.107 & 0.981 & 0.830 & 0.617  \\
$ 5500 $ &  1  & 0.930 & 0.845 & 0.727 & 1.103 & 0.978 & 0.828 & 0.617  \\
\hline
\end{tabular}
\caption{The scaled variances of the primordial and final
positively charged hadrons in central A+A collisions are presented
at different c.m. energies $\sqrt{s_{NN}}$ (see also Fig.~1, left). 
The calculations are
done for the HG with Boltzmann statistics in the GCE. The excluded
volume parameter, $v=16\pi r^3/3$, is taken to be the same for all
hadron species ($r=0$ corresponds to the ideal HG results). The
condition  $n_{tot}\cong1/v$ corresponds to largest possible VDW
total particle density, and this gives an upper limit of the VDW
suppression effect for the scaled variances. \label{Table2}}
\end{center}
\end{table}

\begin{table}[h!]
\begin{center}
\begin{tabular}{||c||c|c|c|c||c|c|c|c||}\hline
 $\sqrt{s_{NN}}$& \multicolumn{4}{c||}{ $\;\omega_{-}$~~~Primordial} & \multicolumn{4}{c||}{
  $\;\omega_{-}$~~~Final} \\ [0.5ex]
\hline [ GeV ] &$r=0$ & $r=\!0.3$fm &
$r=\!0.5$fm&$n_{tot}\cong v^{-1}$ &$~r=0~$ & $r=\!0.3$fm & $r=\!0.5$fm & $n_{tot}\cong v^{-1}$ \\
[0.5ex] \hline\hline
$ 2.32 $ &  1  & 0.997 & 0.989 & 0.946 & 1.001 & 0.998 & 0.988 & 0.941 \\
$ 4.86 $ &  1  & 0.976 & 0.930 & 0.815 & 1.016 & 0.984 & 0.920 & 0.763 \\
$ 6.27 $ &  1  & 0.968 & 0.912 & 0.793 & 1.026 & 0.980 & 0.898 & 0.723 \\
$ 7.62 $ &  1  & 0.963 & 0.901 & 0.782 & 1.035 & 0.977 & 0.883 & 0.700 \\
$ 8.77 $ &  1  & 0.957 & 0.894 & 0.775 & 1.041 & 0.976 & 0.873 & 0.686 \\
$ 12.3 $ &  1  & 0.951 & 0.880 & 0.763 & 1.056 & 0.973 & 0.855 & 0.660 \\
$ 17.3 $ &  1  & 0.944 & 0.870 & 0.755 & 1.068 & 0.972 & 0.843 & 0.644 \\
$ 62.4 $ &  1  & 0.933 & 0.852 & 0.737 & 1.092 & 0.973 & 0.827 & 0.621 \\
$ 130  $ &  1  & 0.931 & 0.849 & 0.732 & 1.097 & 0.975 & 0.827 & 0.619 \\
$ 200  $ &  1  & 0.931 & 0.848 & 0.731 & 1.099 & 0.976 & 0.827 & 0.618 \\
$ 5500 $ &  1  & 0.930 & 0.846 & 0.727 & 1.103 & 0.978 & 0.828 & 0.617 \\
\hline
\end{tabular}
\caption{The same as in Table II, but for negatively charged
hadrons (see also Fig.~2, right).} \label{Table3}
\end{center}
\end{table}

\begin{figure}[ht!]
\begin{center}
\epsfig{file=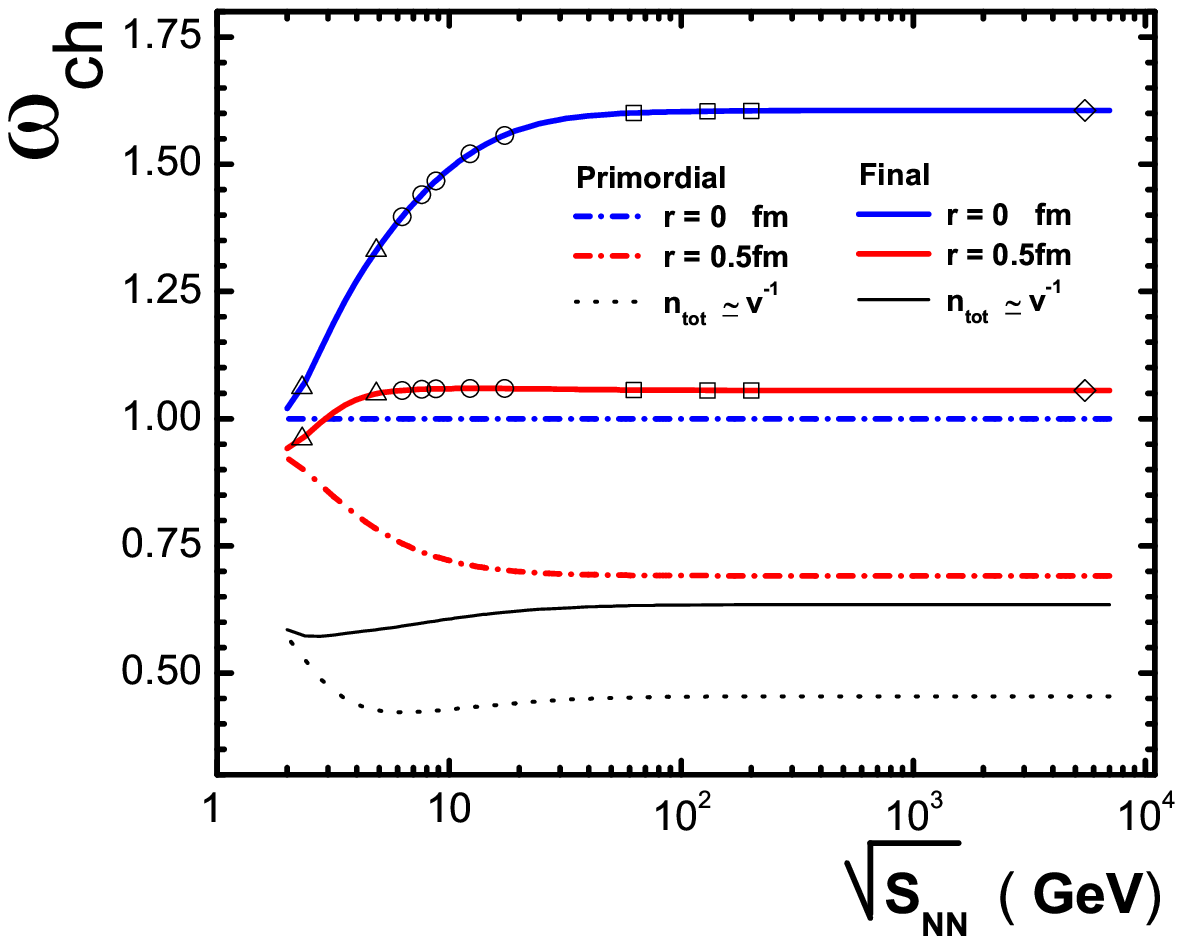,height=6.7cm}
\epsfig{file=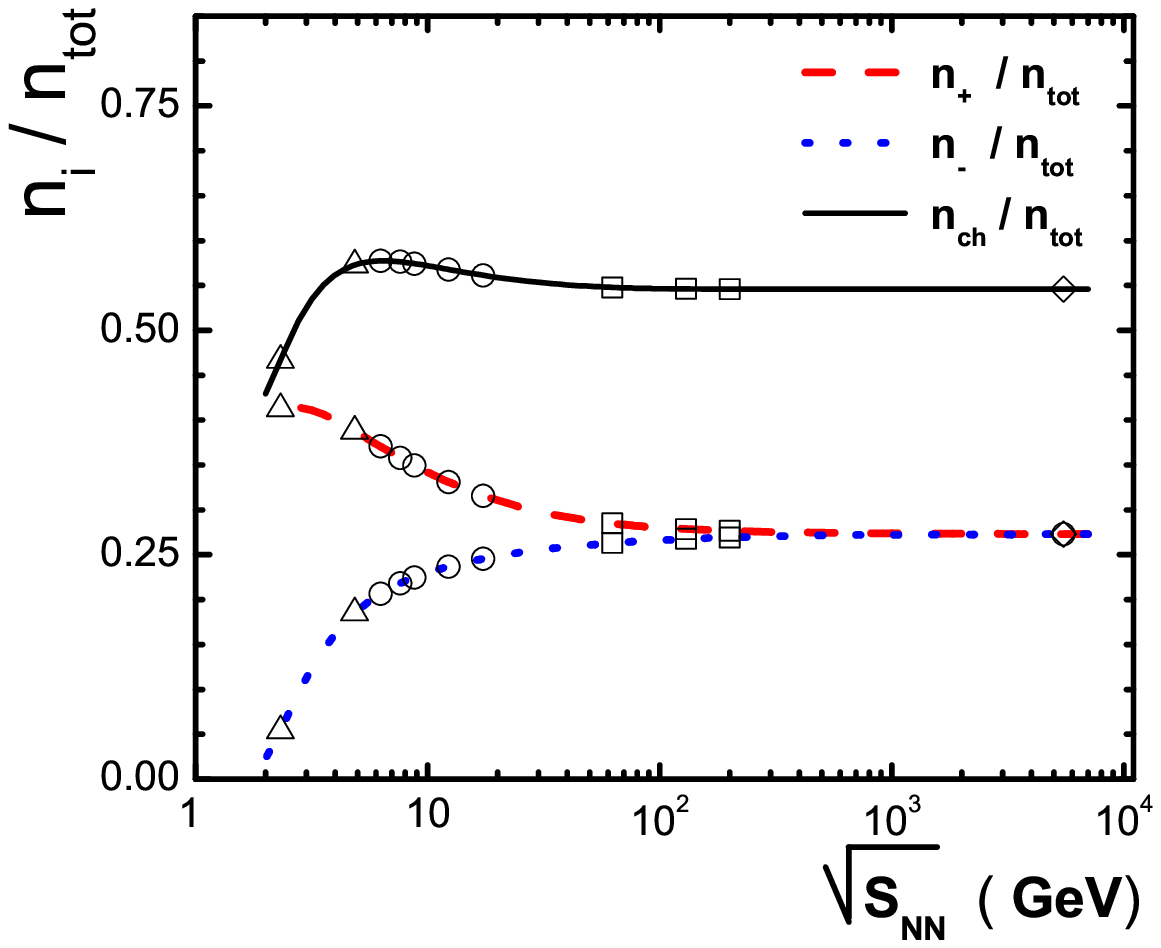,height=6.7cm}
  \caption{{\it Left.} The same as in Fig.~1, but for $\omega_{ch}$ (see also Table IV).
 {\it Right.} The ratios $n_i/n_{tot}=n_i^{id}/n_{tot}^{id}$ (see also Table I) 
 for $i=+$ (dashed line),
 $i=-$ (dotted line), and $i=ch$ (solid line).} \label{omega_ch}
\end{center}
\end{figure}

\begin{table}[h!]
\begin{center}
\begin{tabular}{||c||c|c|c|c||c|c|c|c||}\hline
 $\sqrt{s_{NN}}$& \multicolumn{4}{c||}{ $\;\omega_{ch}$~~~Primordial} & \multicolumn{4}{c||}{
  $\;\omega_{ch}$~~~Final} \\ [0.5ex]
\hline [ GeV ] &$r=0$ & $r=\!0.3$fm &
$r=\!0.5$fm&$n_{tot}\cong v^{-1}$ &$~r=0~$ & $r=\!0.3$fm & $r=\!0.5$fm & $n_{tot}\cong v^{-1}$ \\
[0.5ex] \hline\hline
$ 2.32 $ &  1  & 0.974 & 0.904 & 0.533 & 1.061 & 1.034 & 0.960 & 0.574 \\
$ 4.86 $ &  1  & 0.927 & 0.783 & 0.427 & 1.331 & 1.236 & 1.049 & 0.585 \\
$ 6.27 $ &  1  & 0.912 & 0.755 & 0.423 & 1.397 & 1.274 & 1.056 & 0.592 \\
$ 7.62 $ &  1  & 0.901 & 0.738 & 0.424 & 1.440 & 1.296 & 1.058 & 0.598 \\
$ 8.77 $ &  1  & 0.895 & 0.728 & 0.426 & 1.468 & 1.309 & 1.059 & 0.602 \\
$ 12.3 $ &  1  & 0.882 & 0.712 & 0.432 & 1.521 & 1.331 & 1.059 & 0.612 \\
$ 17.3 $ &  1  & 0.873 & 0.702 & 0.439 & 1.557 & 1.344 & 1.059 & 0.619 \\
$ 62.4 $ &  1  & 0.861 & 0.692 & 0.452 & 1.601 & 1.355 & 1.056 & 0.632 \\
$ 130  $ &  1  & 0.860 & 0.691 & 0.454 & 1.604 & 1.356 & 1.056 & 0.634 \\
$ 200  $ &  1  & 0.860 & 0.691 & 0.454 & 1.605 & 1.356 & 1.056 & 0.634 \\
$ 5500 $ &  1  & 0.860 & 0.691 & 0.454 & 1.605 & 1.356 & 1.056 & 0.634 \\
\hline
\end{tabular}
\caption{The same as in Table II, but for all charged hadrons (see also Fig.~2, left).
\label{Table4}}
\end{center}
\end{table}

The excluded volume corrections, i.e. the factors $ \exp \left( -
vp/T \right)$ and $\left( 1+ v\sum_i x_i \right)^{-1}$, are
calculated using the THERMUS package \cite{Thermus}. Numerical
optimization functions allow to solve the transcendental equation
(\ref{pvdw}) for the pressure and, thus, find the hadron yields
and fluctuations of the VDW HG. The thermodynamical parameters
$T,~\mu_B$, and $\gamma_S$,  the VDW suppression factor
$R=\exp(-vp/T)[1+v\sum_ix_i]$, and the particle number ratios
$n_i/n_{tot}=n_i^{id}/n_{tot}^{id}$ (see also Fig.~2, right) 
along the chemical freeze-out line
are presented in Table I. Once a suitable set of thermodynamical
parameters is determined for each collision energy, the scaled
variance of negatively, positively, and all charged particles can
be calculated using Eq.~(\ref{omega-plus-VDW}) for $\omega_+$ and
similar equations for $\omega_-$, $\omega_{ch}$, with the
correlators taken from Eq.~(\ref{corr-VDW}).   The resulting
scaled variances are presented in Tables II-IV and Figs.~1 and 2 (left)
as a function of $\sqrt{s_{NN}}$. The values of $\sqrt{s_{NN}}$ quoted
in Tables~I-IV and marked in Figs.~1-2 correspond to the beam
energies at SIS (2$A$~GeV), AGS (11.6$A$~GeV), SPS ($20A$, $30A$,
$40A$, $80A$, and $158A$~GeV), colliding energies at RHIC
($\sqrt{s_{NN}}=62.4$~GeV, $130$~GeV, and $200$~GeV), and LHC
($\sqrt{s_{NN}}=5500$~GeV). Both the primordial and final state
scaled variances are presented. To make a correspondence with real
measurements, both strong and electromagnetic decays should be
taken into account, while weak decays should be omitted.

Some features of the results should be mentioned. The values
$\omega_{\pm}=\omega_{ch}=1$ are shown by the dashed-dotted lines
in Figs.~1-2. They correspond to the primordial fluctuations in
the ideal HG ($r=0$). The bump in $\omega_+$ for final state
particles seen in Fig.~\ref{omega_pos} at the small collision
energies  is due to a correlated production of proton and $\pi^+$
meson from $\Delta^{++}$ decays. This single resonance
contribution dominates in $\omega_+$ at small collision energies
(small temperatures), but becomes relatively unimportant at the
high collision energies.

The number of  negative particles is relatively small, $\langle
N_-\rangle \ll \langle N_+\rangle$, at low collision energies (see
Fig.~2, right). For the hard-core radius $r=0.5$~fm, this region
corresponds 'small' particle number density, $vn_{tot}\ll 1$. It
then follows, for the primordial values, $\omega_{\pm}\cong 1- 2v
n_{\pm}$ and $\omega_{ch}\cong 1- 2v n_{ch}$. Thus, the VDW
suppression effects are seen at low collision energy for
positively charged and all charged particles (Fig.~1 and 2), and
are absent for negatively charged particles (Fig.~1, right).

At high collision  energies (large $T$) one finds $n_-\cong n_+$
as seen from Fig.~2, right. For the hard-core radius $r=0.5$~fm,
Fig.~1 shows the asymptotic values of 0.85  for the primordial and
0.83 for the final scaled variances $\omega_{\pm}$ in the VDW HG,
instead of 1 and 1.1, respectively, in the ideal HG. Thus, one
observes about 15\% and 25\% VDW excluded volume suppression for,
respectively, the primordial and final values of $\omega_{\pm}$.
The upper limit of the VDW suppression effects for the scaled
variances is obtained at $n_{tot}\cong 1/v$ and is presented in
Figs.~1-2. The approximate relations both at small density,
$\omega_j\cong 1-2vn_j$, and high density, $\omega_j\cong 1
-n_j/n_{tot}$, demonstrate that VDW suppression of $j$th
fluctuations is proportional to $j$th density in the HG. This
explains why the VDW suppression for the primordial $\omega_{ch}$
seen in Fig.~2 is approximately the same as for $\omega_+$ at
small collision energy and 2 times larger than that for
$\omega_{\pm}$ at high collision energies.

A comparison of the  primordial scaled variances with those for
final hadrons demonstrates that the fluctuations increase in GCE
ideal HG for each hadron species (see Eq.~(\ref{omega-id-1})) for
positive, as well as for negative, and all charged hadrons (see
Figs.~1-2). In the VDW HG the behavior is more complicated. At
high collision energies, as demonstrated in Fig.~1, they even
decrease for $\omega_{\pm}$ after resonance decays. This
unexpected behavior follows from Eqs.~(\ref{VDW-corr}) and
(\ref{corr-VDW}). Calculating  (\ref{corr-VDW}) one finds that
only one term, $\sum_R\langle N_R\rangle \langle n_in_j\rangle$,
coming from resonance decays gives a positive contribution to the
$\langle \Delta N_j \Delta N_k\rangle$ correlator. Other terms in
the right hand side of Eq.~(\ref{corr-VDW}) appear due to
particle-resonance and resonance-resonance (anti)correlations in
the VDW HG. As seen from Eq.~(\ref{VDW-corr}) these terms are
negative and, thus, suppress the scaled variances for the final
state particles. The resulting effect from resonance decays is
defined by the competition of two effects: decays into more than
one $j$th particle against particle-resonance and
resonance-resonance anticorrelations. Multi-particle decays lead to
increase the fluctuations, while the anticorrelations because of
the excluded volume lead to decrease them. For all charged
particles the multi-particle decays win and $\omega_{ch}$ always
increases due to resonance decays as seen from Fig.~2. This same
is true for $\omega_+$ at small collision energies. The opposite
situation takes place for $\omega_{\pm}$ at high collision
energies. Anticorrelations because of the excluded volume overcome
the multi-particle decay contributions, and $\omega_{\pm}$ decreases
as seen from Fig. 1.

\section{summary}
The hadron multiplicity fluctuations in relativistic
nucleus-nucleus collisions have been considered in the statistical
hadron-resonance gas model.  We study the effect of the van der
Waals excluded volume on the hadron distribution scaled variances.
In the present paper we restrict ourself to the grand canonical
calculations within the Boltzmann statistics approximation.

If the proper volume parameter is the same for different hadron
species, the particle number ratios remain unchanged and equal to
those  in the ideal hadron-resonance gas.  Therefore, with
rescaling of the total volume $V$ one may compensate the excluded
volume suppression of the hadron densities leaving the hadron
yields the same as in the ideal hadron-resonance gas. In the
present paper it has been demonstrated that the multiplicity
fluctuations are suppressed in the van der Waals gas. This
suppression is qualitatively different from that of the particle
yields. In contrast to the average multiplicities, the suppression
of multiplicity fluctuations can not be removed by rescaling of
the total volume of the system. 

In this work we have considered two `reasonable` values of hard-sphere
radii, $r=0.3fm$ and $r=0.5fm$, as well as the limiting behavior of the VDW
suppression. Estimates of possible excluded volume suppression effects of
particle number fluctuations from existing data in A+A collisions
will be the subject of future studies.

\begin{acknowledgments}
We would like to thank  V.V. Begun, E.L.~Bratkovskaya,
A.I.~Bugrij, M.~Ga\'zdzicki, W.~Greiner, V.P.~Konchakovski
for discussions. One of the author (M.I.G.) is thankful to
the Humboldt Foundation for the financial support.
\end{acknowledgments}

\end{document}